\documentclass[
reprint,
superscriptaddress,
 amsmath,amssymb,
 aps,
 pre,
floatfix,
]{revtex4-2}

\usepackage{lipsum}
\usepackage{color}
\usepackage{natbib}
\usepackage{graphicx}
\usepackage{dcolumn}
\usepackage{bm}

\begin{document}


\title{Air drag controls the runout of small laboratory landslides}

\author{Rory T. Cerbus}
 \email{rory.cerbus@riken.jp}
 \affiliation{RIKEN Center for Biosystems Dynamics Research (BDR), 2-2-3 Minatojima-minamimachi, Chuo-ku, Kobe 650-0047, Japan
}
\affiliation{
Laboratoire Ondes et Mati\`{e}re d'Aquitaine (LOMA), UMR 5798 Université Bordeaux et CNRS, 351 cours de la Lib\'{e}ration, 33405 Talence, France
}

\author{Ludovic Brivady}
\affiliation{
Laboratoire Ondes et Mati\`{e}re d'Aquitaine (LOMA), UMR 5798 Université Bordeaux et CNRS, 351 cours de la Lib\'{e}ration, 33405 Talence, France
}

\author{Thierry Faug}
\affiliation{
Univ. Grenoble Alpes, CNRS, INRAE, IRD, Grenoble INP, IGE, 38000 Grenoble, France
}



\author{Hamid Kellay}
\affiliation{
Laboratoire Ondes et Mati\`{e}re d'Aquitaine (LOMA), UMR 5798 Université Bordeaux et CNRS, 351 cours de la Lib\'{e}ration, 33405 Talence, France
}


\date{\today}

\begin{abstract}
Laboratory granular landslides are smaller-scale, simplified, yet well-controlled versions of larger and often tragic natural landslides. Using systematic experiments and scaling analysis, we quantify the influence of grain size, fall height, and landslide volume on runout distance. We also determine the minimum landslide size required to observe this scaling, which we find is set by a combination of air drag, grain size, and fall height.
\end{abstract}

\maketitle

\section{Introduction}

A common goal of laboratory-scale granular landslide experiments is to shed light on natural landslides. One prominent qualitative feature they must then reproduce is the classical positive correlation between landslide volume $V$ and runout distance $L$, while also taking into account the total fall height $H$. Although laboratory experiments are able to reproduce the same qualitative behavior \cite{lucas2014frictional,davies1999runout}, up-scaling the results found in the laboratory to what is observed in nature presents a serious challenge \cite{iverson1989dynamic,delannay2017granular,kesseler2020grain}. Thus particular attention is paid to so-called scale effects, wherein variables such as the non-dimensional stresses presumably retain a size dependence even after non-dimensionalization of all relevant variables and matching of key parameters such as Froude or Reynolds numbers \cite{iverson2015scaling}. Notwithstanding these obstacles, we have elsewhere found that by additionally accounting for the grain size, we are able to bring the normalized runout distance of dense laboratory flows into quantitative agreement with the runout of a variety of natural flows through a new scaling of $L$ with $H$, the volume $V$, and the size of the grains $D$ \cite{companionPRL}. This suggests that laboratory-scale experiments can be fruitfully used to systematically investigate not only the qualitative but also the quantitative behavior of even large-scale landslides thousands of times larger. Here we further probe the limitations of this scaling, making estimates of the parameter ranges necessary to observe this scaling in laboratory granular landslides.

Following previous work \cite{iverson2015scaling}, we focus on stresses and in particular the role played by the viscous stress due to the surrounding air. Börzsönyi et al. demonstrated that air drag can significantly affect the phase and flow behavior of strongly inclined laboratory granular flows \cite{borzsonyi2006rapid}. Kesseler et al. performed a systematic experimental and numerical study of air drag on the runout of laboratory granular flows, using a grain Reynolds number to encapsulate its role as a scale effect \cite{kesseler2020grain}. Here we attempt a similar approach but use instead the framework supplied by the aforementioned scaling law to which both dense laboratory granular flows and natural flows conform \cite{companionPRL}. We postulate a gas-liquid granular transition controlled by air drag and gravity to explain deviations from this scaling and determine the landslide parameters which set this transition point. To ascertain this transition point we focus on the scaling behavior of the landslide front speed $U$, which we connect to the runout $L$ by a simple estimate. 

\section{Experimental Methods}
\label{sec:exptMethods}




\begin{figure*}
\centering
\includegraphics[width=1.0\linewidth]{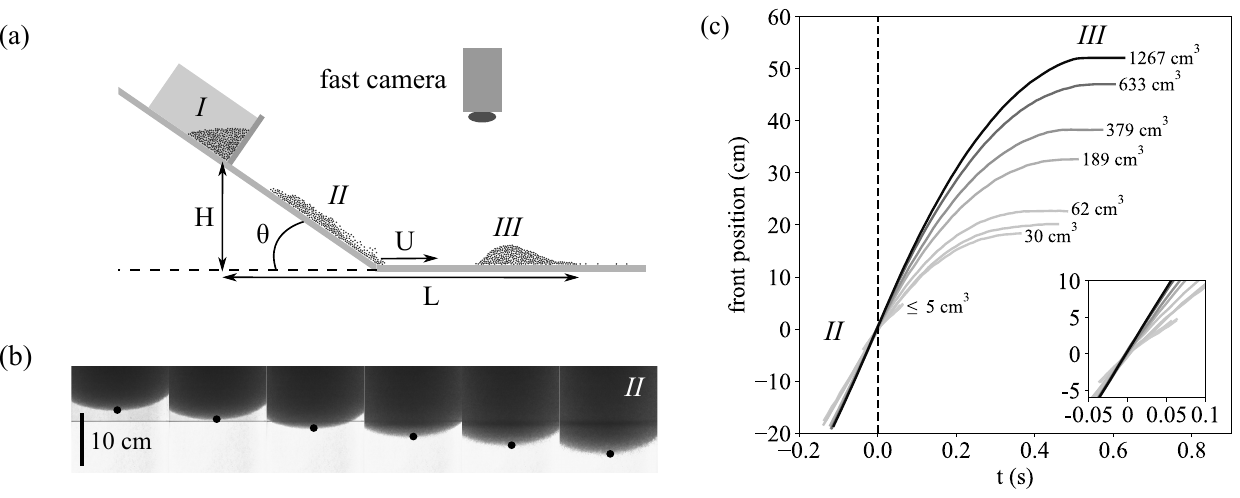}
\caption{(a) Side-view schematic of the experimental setup. The grains are released from a rectangular box ($I$) of width 15 cm via a sliding sluice gate at the front and slide down a flat glass plate (80 cm long and 65 cm wide), inclined at an angle $\theta \simeq 34^\circ$, and eventually reach the junction ($II$) before coming to rest on a level flat glass plate (125 cm by 125 cm). The motion of the grains is observed with an overhead fast camera to determine the front speed $U$ at the junction. (b) Representative image series of an experimental granular landslide near the junction (horizontal line in all images) at intermediate times ($II$). Time is from left to right and the flow direction is down. Images are 0.01 s apart. The black dot identifies the front position. Here the total volume is $V \simeq 633$ cm$^3$ ($M \simeq 1000$ g), and the grain size is $\langle D \rangle = 562 \mu$m. (c) Time series of front position for several different volumes for the grain size $\langle D \rangle = 328 \mu$m. The time $t=0$ is defined as the moment when the front crosses the junction. Both the slope, from which we determine $U$, and the final distance depend on $V$. Inset: Zoom-in near the junction crossing. Below $V \lesssim 5$ cm$^3$ the slope is nearly constant.}
\vspace{-0em}
\label{setup}
\end{figure*}





We begin by describing our laboratory experiments for which we systematically vary the fall height $H$, the total volume $V$, and the grain size $D$. We have found that the grain size distribution is a key element to understanding the runout of granular landslides, in particular through $\langle D \rangle$, $\langle D^3 \rangle$, and their ratio $S = \langle D^3 \rangle / \langle D \rangle ^3$. Our main observables are the runout distance $L$, and the landslide front speed $U$. The experimental setup is a simplified version of a natural landslide consisting of a slope, a flat section, grains, and a container to house the grains before releasing them by rapidly raising a sliding metal gate. See Fig.~\ref{setup}a for a schematic of the experimental setup.

\begin{figure*}[t!]
\centering
\includegraphics[width=1.0\linewidth]{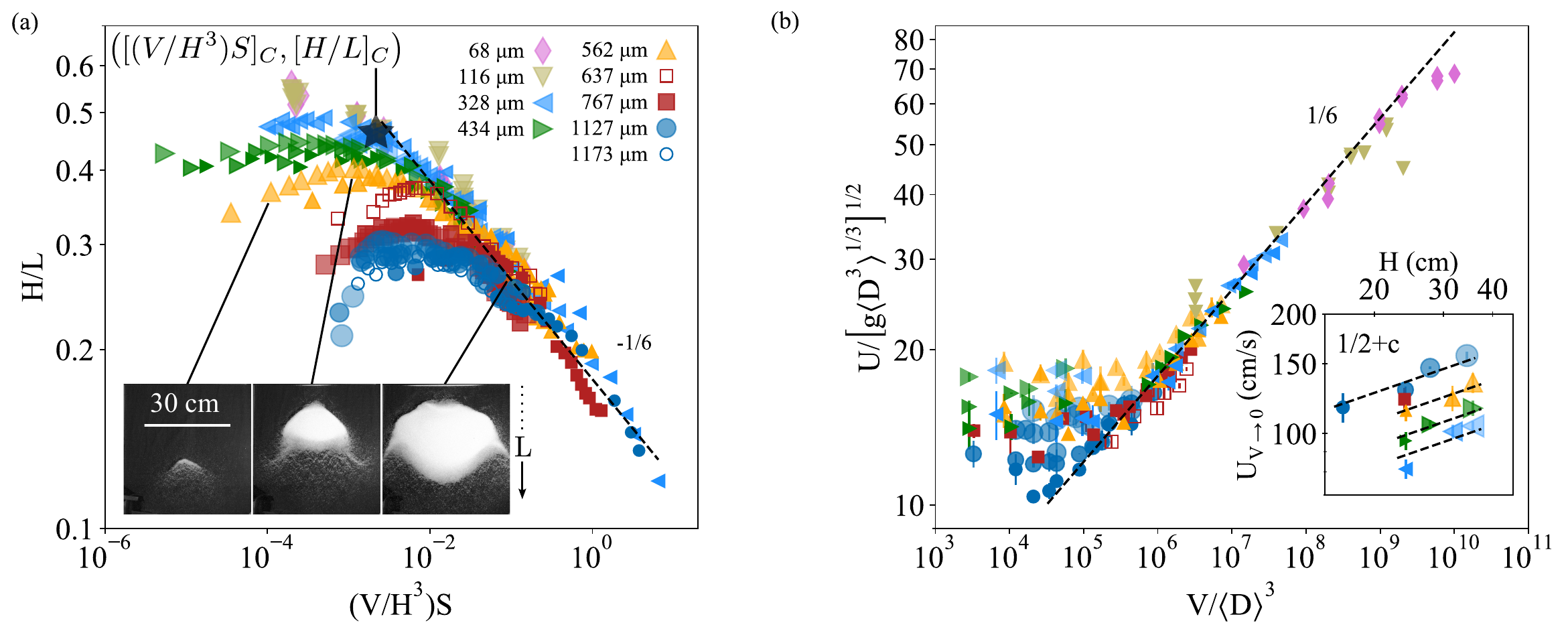}
\vspace{-2em}
\caption{(a) Normalized runout distance $H/L$ vs. $(V/H^3)S$ for different values of $\langle D \rangle$ and $H$ (data symbols with larger $H$ are more transparent and bigger). At large $(V/H^3)S$, all data conform to a master curve $\sim \left[ (V/H^3)S \right]^{-1/6}$ which we identify with the dense, liquid regime. At low $(V/H^3)S$, the curves peel off the master curve according to their $\langle D \rangle$ and $H$. The departure from the master curve occurs horizontally at $\left[ (V/H^3)S \right]_C$ and vertically at $[H/L]_C$. We point out one such example with a dark, transparent star. The departure and the maxima of each curve appear to depend on $\langle D \rangle$ and $H$ in a similar way. Inset: Overhead images of the final landslide position for low, intermediate, and large $N$ for $H$ = 31.46 cm, $\langle D \rangle$ = 562 $\mu$m. The upper part of each image corresponds to the junction between the inclined and flat sections and the runout direction is oriented downwards. (b) Plot of normalized front speeds $U/\sqrt{g \langle D^3 \rangle^{1/3}}$ vs. $V/\langle D \rangle^3$ for laboratory experiments at various $\langle D \rangle$ and $H$ (symbols as in (a)). At large $V$, $U/\sqrt{g \langle D^3 \rangle^{1/3}} \sim (V/\langle D \rangle ^3)^{1/6}$. At low $V$ the $U$ data peel off the $(V/\langle D \rangle ^3)^{1/6}$ curve at a value of $(V/\langle D \rangle ^3)$ determined by $H$ and $\langle D \rangle$ (see Sec.~\ref{sec:diluteLimit}). A fit of $(V/\langle D \rangle ^3)^{1/6}$ to high $(V/\langle D \rangle ^3)$ experimental data ($--$) yields $R^2 \simeq 
0.95$. Inset: A plot of $U_{V \rightarrow 0}$ vs. $H$ for different $\langle D \rangle$ (same symbols as main figure) reveals a strong dependence on $\langle D \rangle$. Likewise the slope of the line $1/2+c$ is smaller ($c \simeq -0.1576$) than the 1/2 expected when only gravity is relevant, suggesting additional significant stresses.}
\label{runoutResult}
\end{figure*}

For each experiment we measure the total mass $M$ with a simple scale and gently sprinkle the grains into the rectangular box so that the surface is level. We measure $V$ (in the box), $H$, and $L$ using a standard measuring tape with millimeter precision. Because individual grains can escape and travel farther than others, we identify the final front position as the frontmost position where a layer of grains is still in contact with the main mass (see Fig.~\ref{setup}a and Ref.~\cite{companionPRL}). In addition, we also measure the landslide front position and speed $U$ using an overhead fast camera (Phantom v641) at a frame rate of 100 Hz. Our grains are relatively spherical glass beads which have been roughly pre-sorted by the manufacturer according to diameter $D$, ranging from $\sim$ 45 $\mu$m to $\sim$ 1.5 mm (see App.~\ref{sec:labGrains}). We characterize each collection of grains by measuring their size distributions using an imaging technique (see App.~\ref{sec:labGrains}). This yields the frequency distribution $p_f(D)$, which we then convert into the mass-weighted distribution $p(D) \propto p_f(D) \times (4 \pi \rho/3) (D/2)^3$, where $\rho$ is the grain material mass density. We calculate average quantities such as the average diameter $\langle D \rangle$ using $p(D)$. 
While electrostatic effects can be prominent for very small particle sizes  \cite{shinbrot2012electrostatic}, the present experiments were performed for a range of room humidities and temperatures and a range of different particle sizes. 
The collapse which we find (see Fig.~\ref{runoutResult}) is without reference to such effects.



To determine the front speed $U$ in the laboratory experiments, we used a standard image processing tool (ImageJ) to manually track the front position of the landslide. As shown in Fig.~\ref{setup}b, a series of zoomed-in images of near the junction between the inclined and flat section of the experiments, the front is easily distinguished (represented by a black dot). We fit a line to the front positions for the first three time steps after the junction. We estimate $U$ as the slope of this line, and thus $U$ is really an initial front speed which is necessarily followed by deceleration after the junction. We estimated an error bar in this determination of $U$ by the difference between the speed estimated by fitting three or four points. 

\section{Preliminary Observations}

\subsection{Runout and front speed}

Elsewhere we considered only experiments and natural field data at large $V$ for which the flow is dense \cite{companionPRL}. As shown in Fig.~\ref{runoutResult}a, all of the experimental runout data for different $H$ and $\langle D \rangle$ at large $V$ all conform to the following scaling:
\begin{equation}
    \frac{H}{L} \sim \left[ \left( \frac{V}{H^3} \right) S \right]^{-1/6}.
    \label{eq:mainResultLiquid}
\end{equation}
Here we consider a wider range of flow densities by reducing $V$ by several orders of magnitude. While at large densities, in correspondence with natural flows, there is a positive correlation between the runout distance $L$ and $V$, we find different behavior at small $V$. This manifests as a non-monotonic \cite{cerbus2021landslide} deviation from the scaling (Eq.~(\ref{eq:mainResultLiquid})) which can be seen in Fig.~\ref{runoutResult}a. At low $V$, $H/L$ increases with $V$, opposite to the classical trend \cite{legros2002mobility}. Moreover the point of departure appears to depend on both $D$ and $H$. Here we take as our main objective determining the point of deviation from Eq.~(\ref{eq:mainResultLiquid}) and its dependence on typical landslide parameters. This will also serve to determine the range of validity of the large $(V/H^3)S$ scaling in Eq.~(\ref{eq:mainResultLiquid}).

Parallel to the scaling of the runout distance $L$ in Fig.~\ref{runoutResult}a is the scaling of the front speed $U$ shown in Fig.~\ref{runoutResult}b. In the large $V$ limit \footnote{For simplicity we represent the limits in $(V/H^3)S$ or $V/\langle D \rangle^3$ as limits in $V$, since $H$ and $D$ are always fixed for a set of experiments.}, $U_{V \rightarrow \infty}/\sqrt{g \langle D^3 \rangle^{1/3}}$ becomes independent of $\langle D \rangle$ and $H$ and depends only on $(V/\langle D \rangle^3)$. We find empirically that:
\begin{equation}
    \frac{U_{V \rightarrow \infty}}{\sqrt{g \langle D^3 \rangle^{1/3}}} \sim (V/\langle D \rangle^3)^{1/6}.
    \label{eq:scalingUlargeN}
\end{equation}
This scaling is consistent with the scaling of the front speed found for a large avalanche of ping-pong balls on a ski jump \cite{mcelwaine2001ping}, and the power-law exponents in Eq.~(\ref{eq:mainResultLiquid}) and Eq.~(\ref{eq:scalingUlargeN}) immediately suggest a connection between the runout distance $L$ and the initial front speed $U$. On the other hand, the front speeds in the low $V/\langle D \rangle^3$ limit, $U_{V \rightarrow 0}$, become independent of $V/\langle D \rangle^3$ and retain a dependence on both $H$ and $\langle D \rangle$ (see App.~\ref{sec:airDrag}). 

The non-monotonic behavior in $L$ can easily be observed in images of the final grain pile (see inset to Fig.~\ref{runoutResult}a). As these images qualitatively suggest, the region of decrease corresponds to a dense flow with many inter-particle interactions, and the region of increase corresponds to a dilute flow with fewer interactions. We thus postulate that the transition between the regimes of increasing and decreasing normalized runout are manifestations of an overall gas-liquid granular transition \cite{borzsonyi2006rapid}, and we refer to these two regimes as gaseous (dilute) and liquid (dense). We will test this hypothesis by determining the scaling of the intersection of the gaseous and liquid regimes, roughly corresponding to the maxima denoted by $\left[ (V/H^3)S \right]_C$ and $[H/L]_C$ in Fig.~\ref{runoutResult}a, and comparing this to our experimental data. We will take advantage of the close connection between $L$ and $U$ in both the large and small $(V/H^3)S$ regimes to estimate at which point the dense scaling (Eq.~(\ref{eq:mainResultLiquid})) fails.

\subsection{Breakdown of landslide scaling: the dilute limit}
\label{sec:diluteLimit}

Here we consider the dilute limit. It is natural to assume that inter-grain forces, which are apparently important in the dense limit $V \rightarrow \infty$, have become irrelevant, as reflected by the independence from $(V/\langle D \rangle^3)$. A dependence on $H$ as seen in Fig.~\ref{runoutResult}b is natural and would follow from considering gravity only, but instead of the expected dependence $U_{V \rightarrow 0} \sim H^{1/2}$, we find $U_{V \rightarrow 0} \sim H^{1/2 + c}$, with $c < 0$. Following Refs. \cite{borzsonyi2006rapid,kesseler2020grain} we postulate the role of air drag, modeled as the drag on a single sphere \cite{schlichting2016boundary,tsuji1982fluid,haider1989drag}, in addition to gravity. With these minimal ingredients we perform a scaling analysis for $U_{V \rightarrow 0} = F \left(g,H,\langle D \rangle,\nu_{\rm{air}},\rho_{\rm{air}},\rho_{\rm{P}} \right)$, where including $\nu_{\rm{air}}$ and $\rho_{\rm{air}}$ anticipates the role of air drag.

\begin{figure}
\centering
\includegraphics[width=1\linewidth]{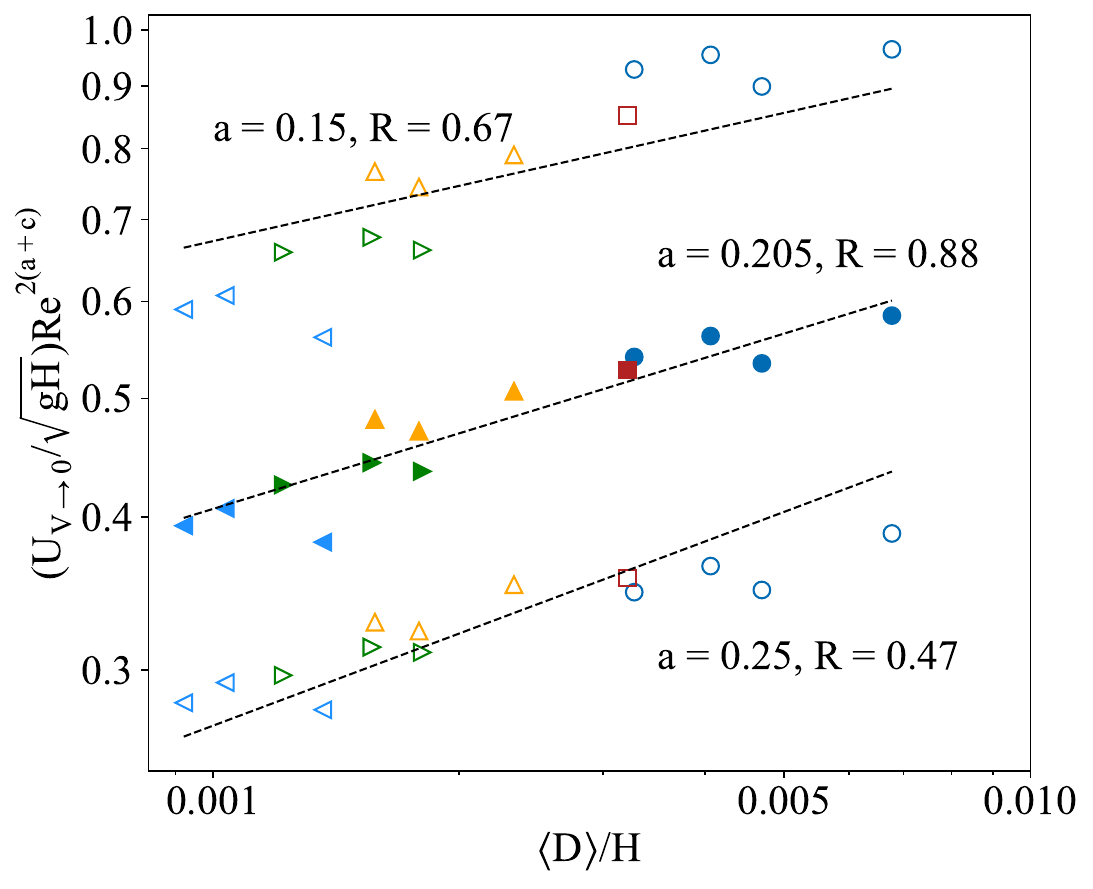}
\vspace{-2em}
\caption{Plot of $U_{V \rightarrow 0} / \sqrt{gH} Re^{2(a+c)}$ vs. $\langle D \rangle / H$ used to determine the value of the exponent $a$ empirically. In accord with Eq.~(\ref{eq:UN0scaling}), $U_{V \rightarrow 0} / \sqrt{gH} Re^{2(a+c)}$ should be proportional to $(\langle D \rangle / H)^a$, so that in this log-log plot the slope of the line should be $a$. Adjusting $a$ as a fitting parameter shows that a value of $a \simeq$ 0.205 (solid symbols) yields a good fit, whereas slightly different values of $a$ do not (open symbols). The goodness of fit for $a \simeq 0.205$ also justifies the scaling predicted in Eq.~(\ref{eq:UN0scaling}).}
\label{determiningA}
\end{figure}

\begin{figure*}[t!]
\centering
\includegraphics[width=1\linewidth]{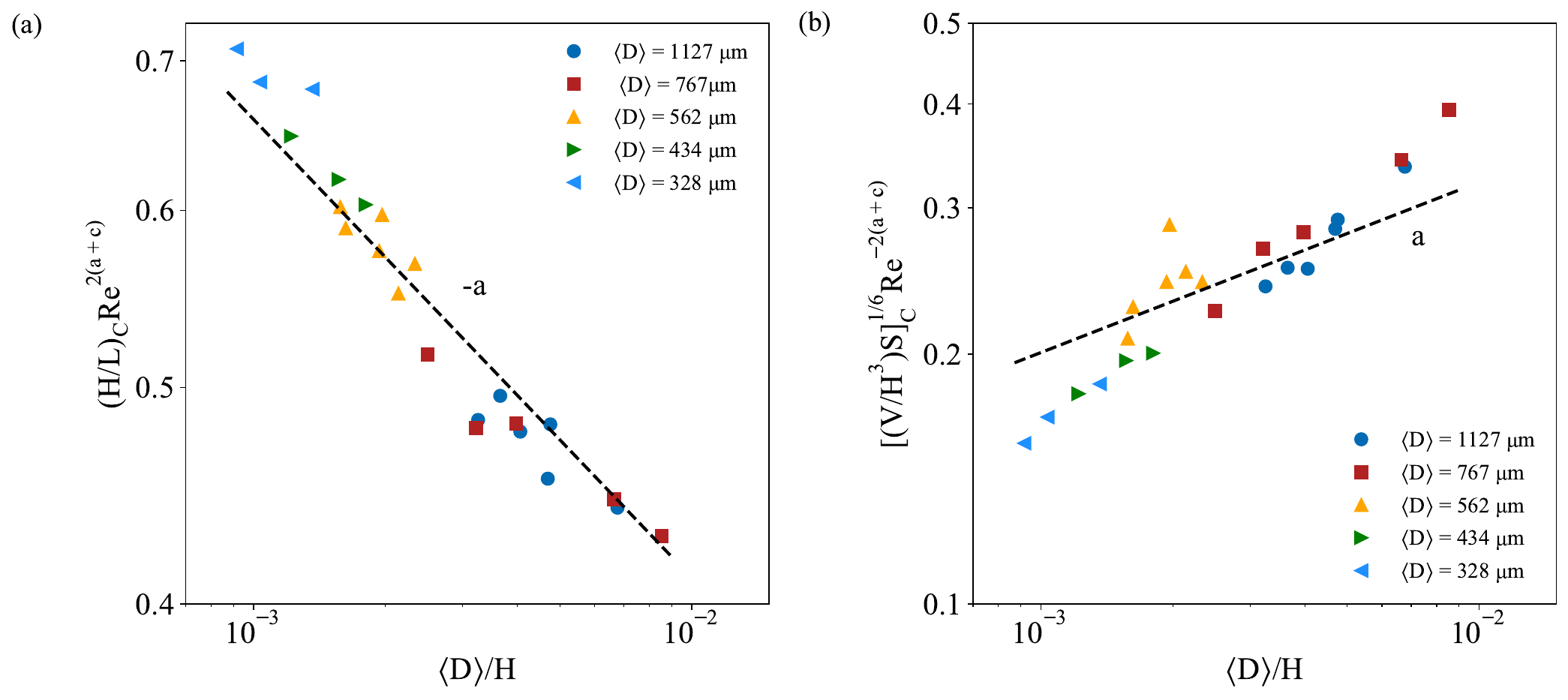}
\vspace{-2em}
\caption{Plots of the critical values of the effective friction ($H/L$) and $(V/H^3)S$, newly normalized according to the predicted scalings (Eqs.~\ref{eq:Nmax},\ref{eq:HLmax}). (a) Log-log plot of $(H/L)_C Re^{2(a+c)}$ vs. $\langle D \rangle / H$. According to Eq.~(\ref{eq:HLmax}) the slope of this curve should be $-a$, which agrees well with the data. (b) Log-log plot of $[(V/H^3)S]_C^{1/6} Re^{-2(a+c)}$ vs. $\langle D \rangle / H$. According to Eq.~(\ref{eq:Nmax}) the slope of this curve should be $a$, which agrees satisfactorily with the data.}
\label{finalResult}
\end{figure*}

Performing a standard dimensional analysis (see App.~\ref{sec:airDrag} for details), we assume a power-law dependence on the ratio $\langle D \rangle / H$ and introduce a grain size-dependent Reynolds number $Re = \sqrt{g H} \langle D \rangle / \nu_{\rm{air}}$,
\begin{equation}
    \frac{U_{V \rightarrow 0}}{\sqrt{g H}} \sim \left(\frac{\langle D \rangle}{H} \right)^a Re^{2(a+c)}.
    \label{eq:UN0scaling}
\end{equation}
The exponent $c \simeq$ -0.1576 $\pm 0.071$ is determined empirically from the data in the inset to Fig.~\ref{runoutResult}b. To determine an appropriate value of $a$ we plot in Fig.~\ref{determiningA} the normalized values of $U_{V \rightarrow 0} / \sqrt{gH} Re^{2(a+c)}$ vs. $\langle D \rangle / H$. If the scaling in Eq.~(\ref{eq:UN0scaling}) is correct, then $U_{V \rightarrow 0} / \sqrt{gH} Re^{2(a+c)} \sim (\langle D \rangle /H)^a$, and so the data, in this log-log plot, will appear as a straight line with slope $a$. The best fit for $a$ determined in this way yields $a \simeq$ 0.205 $\pm 0.05$ (solid symbols in Fig.~\ref{determiningA}). Note that values of $c = 0$ and $a = 0$ correspond to the solution for a single grain descending a height $H$ in the absence of air drag.
 
We now equate the scaling of the front speed for the gaseous regime, $U_{V \rightarrow 0}$ (Eq.~(\ref{eq:UN0scaling})), and for the liquid regime, $U_{V \rightarrow \infty}$ (Eq.~(\ref{eq:scalingUlargeN})). This determines the dependence of $[(V/H^3)S]]_C$ and $(H/L)_C$ on $\langle D \rangle$ and $H$ through the parameters $\langle D \rangle / H$ and $Re$. Setting the high and low $V$ scaling for the front speeds equal,
\begin{align}
\begin{split}
    \left[ \sqrt{g \langle D^3 \rangle^{1/3}} (V/\langle D \rangle^3)^{1/6} \right] _C \sim U_{V \rightarrow \infty} \sim U_{V \rightarrow 0} \\
    \sim \sqrt{gH} \left( \frac{\langle D \rangle}{H} \right)^a Re^{2(a+c)},
\end{split}
\label{eq:equateGasLiquid}
\end{align}
yields 
\begin{equation}
    \left[ (V/H^3)S \right]_C^{1/6} \sim \left( \frac{\langle D \rangle}{H} \right)^{a} Re^{2(a+c)}.
    \label{eq:Nmax}
\end{equation}
The scaling of $[(V/H^3)S]_C$ in turn yields the scaling of $(H/L)_C$. Using the dense liquid result (Fig.~\ref{runoutResult}a) for the normalized runout $(H/L)_C \sim [(V/H^3)S]_C^{-1/6}$, this yields 
\begin{equation}
    \left( \frac{H}{L} \right)_C \sim \left( \frac{\langle D \rangle}{H} \right)^{-a} Re^{-2(a+c)}.
    \label{eq:HLmax}
\end{equation}
We test these predictions in Fig.~\ref{finalResult} by normalizing $(H/L)_C$ and $[(V/H^3)S]_C$ according to Eqs.~(\ref{eq:Nmax}),(\ref{eq:HLmax}). This yields excellent collapse of the newly normalized critical values, the putative intersection of the liquid and gaseous regimes. Starting from the hypothesis that the non-monotonic behavior, and in particular the maxima of the curves in Fig.~\ref{runoutResult}a, is a manifestation of a granular gas-liquid transition, we were able to arrive at a prediction for the scaling of these maxima which agrees well with experiments. Previous studies have argued that air drag in laboratory landslides may prohibit them from being compared to natural landslides where air drag is less relevant due to the much larger $\langle D \rangle$ \cite{kesseler2020grain}. Here we have performed a complete analysis to find the scaling that will determine whether a granular flow is to be considered gaseous or liquid, and thus provide a useful tool for constructing laboratory analogs of natural landslides which are typically in the dense flow regime. If the normalized volume of grains $(V/H^3)S$ be larger than $[(V/H^3)S]_C \sim (\langle D \rangle / H)^{6a} Re^{12(a+c)}$ in Eq.~(\ref{eq:Nmax}), then the system is in the granular dense regime and the excellent agreement between our laboratory experiments in this dense regime and field data \cite{companionPRL} indicate that here the effect of air drag is minimal, thus enabling a direct comparison.


\begin{figure*}[t!]
\centering
\includegraphics[width=1.0\linewidth]{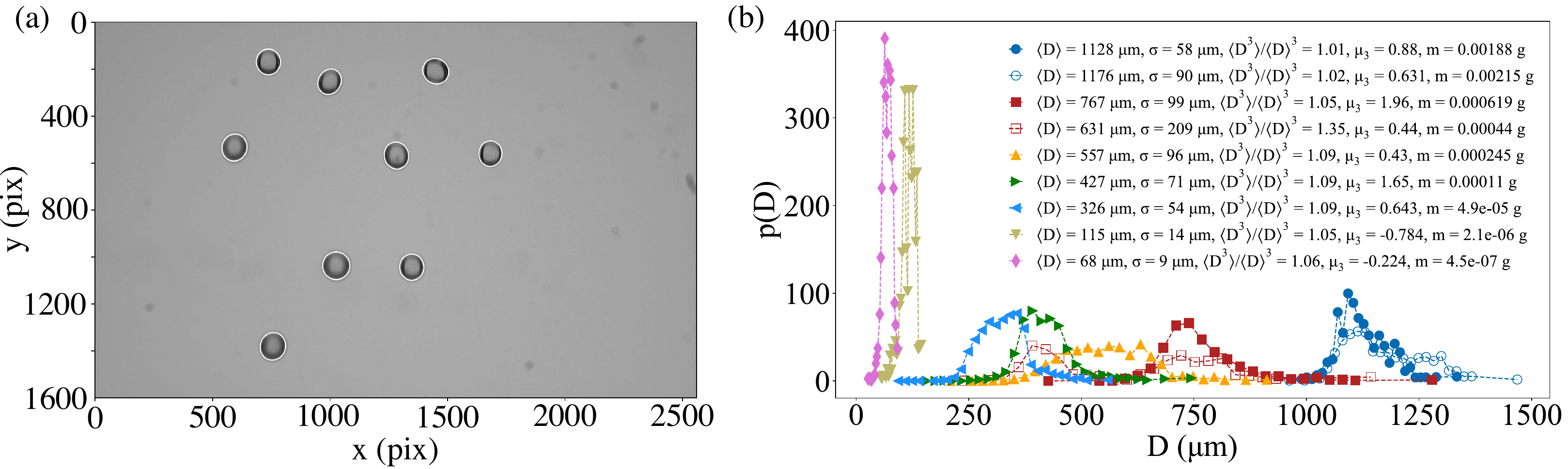}
\vspace{-2em}
\caption{(a) Example particle images used to determine grain size distributions. The particle size is determined by fitting an ellipse to the grains falling through a narrow gap between two glass plates. These grains correspond to the $\langle D \rangle = 1127$ $\mu$m grain distribution in (b). (b) Plot of grain size distribution weighted by mass (or equivalently by volume). The skewness $\langle D^3 \rangle / \langle D \rangle^3$ is also included. Several distributions are slightly bimodal and then sieved to produce another set of grains with a more uniform grain size distribution. The data in Fig.~\ref{runoutResult} include both.}
\label{labGrainSize}
\end{figure*}

\section{Conclusion}

In summary, we have performed an extensive experimental study of laboratory landslide runout combined with a simple scaling analysis to further our understanding of the dominant parameters and phases that control granular landslides. Performing experiments over a wide range of landslide size revealed unexpected non-monotonic behavior: an increase in runout distance with decreasing landslide size in direct contrast to the classical trend. We demonstrated that this is a manifestation of a granular gas-liquid transition. Both gravity and air drag play an important role so that the volume of grains at which the gaseous behavior turns into the universal liquid behavior depends on both $H$ and $\langle D \rangle$ in a predictable way. Analyzing only the dense, granular liquid phase, we found that the runout is controlled by the mass-weighted grain distribution ($\langle D \rangle$ and $\langle D^3 \rangle$), the fall height $H$, and volume of grains $V$. By contrast a semi-empirical analysis of the dilute, granular gaseous phase found that the role of the drag due to the surrounding air needs to be taken into account through the grain Reynolds number, which includes the kinematic viscosity $\nu_{\rm{air}}$. The ability to distinguish whether a laboratory granular landslide is gaseous or liquid is of particular importance as it provides direction in designing experiments to compare with natural landslides.

\emph{Acknowledgments.} R.T.C. gratefully acknowledges funding from the Horizon 2020 program under the Marie Skłodowska-Curie Action Individual Fellowship (MSCAIF) No. 793507, and H.K. acknowledges support from the Institut Universitaire de France. We thank Alexis Petit and Alioune Mbodji for their help in performing some of the experiments. 


\appendix

\section{Grain size distributions}
\label{sec:labGrains}

Using a weight scale and measuring the volume displaced by a large number of grains, we determine the mass density of all laboratory grains (glass grains from OTS) to be $\rho = 2.48 \pm 0.01$ g/cm$^3$. We determine the size distribution of the grains used in our laboratory experiments by imaging the grains (see Fig.~\ref{labGrainSize}a). We sprinkle the grains in between two vertical glass plates and observe them with a steady illumination and fast camera. In this way the particles generally do not touch and we can accurately estimate their size. We fit an ellipse to each grain and defined the diameter $D$ of each grain as the average between the semi-major ($a$) and semi-minor ($b$) axes (see Fig.~\ref{labGrainSize}a). We used our own Python scripts that utilize the OpenCV library \cite{opencv_library} to process the images to determine $D$. This yields the number-weighted probability distribution $p_f(D)$. We confirm this estimate of $p_f(D)$ by counting and measuring the total weight of grains on a sensitive scale (Adam Equipment PW 254) and verify that the average diameter determined from $p_f(D)$, $\int_0^{\infty} p_f(D) dD$, is within 2\% of the estimate from the manual counting method. We then convert the size distributions to the mass-weighted distribution $p(D)$ (see Fig.~\ref{labGrainSize}b). Taking the mass-weighted averages of either grain collection allowed us to satisfactorily collapse all of the data.



\section{Scaling analysis for the dilute limit: further details}
\label{sec:airDrag}


In order to treat the dependence of $U_{V \rightarrow 0}$ on $H$ and $\langle D \rangle$ systematically, we perform a standard dimensional analysis \cite{barenblatt2003scaling}, writing $U_{V \rightarrow 0}$ as an unknown function $F$ given by 
\begin{equation}
    U_{V \rightarrow 0} = F \left( g,H,\langle D \rangle,\nu_{\rm{air}},\rho_{\rm{air}},\rho_{\rm{P}} \right).
\end{equation}
where $\nu_{\rm{Air}}$ and $\rho_{\rm{Air}}$ are the kinematic viscosity and density of air respectively, and $\rho_P$ is the grain density. Buckingham's $\Pi$-theorem \cite{barenblatt2003scaling} enables us to rewrite this general but uninformative equation as a new non-dimensional function of non-dimensionalized variables with fewer arguments. We choose the following non-dimensional groups:
\begin{equation}
    \frac{U_{V \rightarrow 0}}{\sqrt{gH}}, \quad \frac{\rho_{\rm{air}}}{\rho_{\rm{P}}}, \quad \frac{\sqrt{gH} \langle D \rangle}{\nu_{\rm{air}}}, \quad \frac{\langle D \rangle}{H}.
\end{equation}
The first ratio is the dilute front speed $U_{V \rightarrow 0}$ normalized by $\sqrt{gH}$. We choose to normalize the dilute front speed using $H$ rather than $D$ because the overall drop height is a more relevant length scale than the grain size in the absence of inter-particle interactions. The three remaining parameters are the ratio of densities, the ratio of grain size to fall height, and a Reynolds number $Re = \sqrt{gH} \langle D \rangle / \nu_{\rm{air}}$ where the characteristic speed is given by $\sqrt{g H}$, proportional to the value a grain would have at the junction in the absence of air drag. This yields
\begin{equation}
    \frac{U_{V \rightarrow 0}}{\sqrt{gH}} = \Phi \left( \frac{\rho_{\rm{air}}}{\rho_{\rm{P}}}, Re , \frac{\langle D \rangle}{H} \right).
\end{equation}
We now consider whether the function $\Phi$ can be further simplified by separating into three independent functions for each new variable. Because $\rho_{\rm{air}}/\rho_{\rm{P}}$ is constant in our experiments we call this function a constant $C$. Calling the undetermined functions of $\langle D \rangle / H$ and $Re$, $f_{\langle D \rangle / H}$ and $f_{Re}$ respectively, we have $U_{V \rightarrow 0} / \sqrt{gH} = C f_{\langle D \rangle / H} f_{Re}$. The analysis can be simplified even further by making assumptions about $f_{\langle D \rangle / H}$ and $f_{Re}$. Since $\langle D \rangle /H \ll 1$ and $Re \gg 1$, we make the (usual) assumption of incomplete similarity such that $U_{V \rightarrow 0}$ retains its dependence on the small and large parameters in power-law form \cite{barenblatt2003scaling}. This yields $U / \sqrt{gH} \sim (\langle D \rangle / H)^a Re^b$. This will set a constraint on the possible exponents, since $U_{V \rightarrow 0} \sim H^{1/2 + c}$, such that $-a + b/2 = c$ and thus $b = 2(a+c)$, reducing the number of fitting parameters to one, $a$. This yields the main prediction for $U_{V \rightarrow 0}$:
\begin{equation}
    \frac{U_{V \rightarrow 0}}{\sqrt{g H}} \sim \left(\frac{\langle D \rangle}{H}\right)^a Re^{2(a+c)}.
    \label{eq:UN0scalingSupp}
\end{equation}


We test Eq.~(\ref{eq:UN0scalingSupp}) by plotting in Fig.~\ref{determiningA} the normalized speed $(U_{V \rightarrow 0} / \sqrt{gH}) Re^{2(a+c)}$ vs. $\langle D \rangle / H$. The value of $c$ is already determined empirically from the relation $U_{V \rightarrow 0} \sim H^{1/2+c}$ (see inset to Fig.~\ref{runoutResult}b), yielding $c \simeq$ -0.1576 $\pm 0.071$, where the uncertainty is determined by the variation over the different $\langle D \rangle$. We vary the parameter $a$ and find that a value of $a \simeq$ 0.205 yields a good fit to the experimental data, accounting for the dependence on both $H$ and $\langle D \rangle$ (see Fig.~\ref{determiningA}).

\bibliography{landslideFrictionCollapsePRE}

\end{document}